\documentclass[NJP]{iopart}
\usepackage[T1]{fontenc}
\usepackage[latin9]{inputenc}
\usepackage[active]{srcltx}
\usepackage{color}
\usepackage{amsbsy}
\usepackage{amstext}
\usepackage{graphicx}
\usepackage{esint}
\usepackage[numbers,sort&compress]{natbib}
\usepackage[unicode=true,pdfusetitle,
 bookmarks=true,bookmarksnumbered=false,bookmarksopen=false,
 breaklinks=false,pdfborder={0 0 0},pdfborderstyle={},backref=false,colorlinks=true]
 {hyperref}
\hypersetup{
 citecolor=blue,urlcolor=blue}

\makeatletter

\newcommand{\noun}[1]{\textsc{#1}}

\usepackage{iopams}
\usepackage{setstack}

\usepackage{iopams}

\usepackage{bbm}

\makeatother

\begin{document}
\title{Enhancing photoelectric current by nonclassical light }
\author{Hai-Yan Yao, Sheng-Wen Li$^{1}$}
\address{Center for Quantum Technology Research, and Key Laboratory of Advanced
Optoelectronic Quantum Architecture and Measurements, School of Physics,
Beijing Institute of Technology, Beijing 100081, People\textquoteright s
Republic of China}
\ead{$^{1}$lishengwen@bit.edu.cn}
\begin{abstract}
We study the photoelectric current generated by a driving light with
nonclassical photon statistics. Due to the nonclassical input photon
statistics, it is no longer enough to treat the driving light as a
planar wave as in classical physics. We make a  quantum approach to
study such problems, and find that: when the driving light starts
from a coherent state as the initial state, our quantum treatment
well returns the quasi-classical driving description; when the the
driving light is a generic state with a certain \emph{P} function,
the full system dynamics can be reduced as the \emph{P} function average
of many ``branches'' -- in each dynamics branch, the driving light
starts from a coherent state, thus again the system dynamics can be
obtained in the above quasi-classical way. Based on this  quantum
approach, it turns out the different photon statistics does make differences
to the photoelectric current. Among all the classical light states
with the same light intensity, we prove that the input light with
Poisson statistics generates the largest photoelectric current, while
a nonclassical sub-Poisson light could exceed this classical upper
bound.
\end{abstract}
\noindent{\it Keywords\/}: {\emph{Nonclassical light, photon statistics, photoelectric current,
master equation}\\
}
\submitto{\NJP}

\maketitle
\maketitle

\section{Introduction}

When considering a driving light shining on a quantum two-level system
(TLS) ($\hat{H}_{\text{\textsc{s}}}=\hbar\Omega|\mathsf{e}\rangle\langle\mathsf{e}|$,
with $|\mathsf{e}/\mathsf{g}\rangle$ as the excited/ground state),
the interaction between the TLS and the light beam is usually described
by the following \emph{quasi-classical driving} \citep{scully_quantum_1997,gerry_introductory_2005,agarwal_quantum_2012},
\begin{equation}
\hat{V}=-\hat{\boldsymbol{d}}\cdot\vec{E}_{0}\sin(\omega_{\mathbf{k}}t-\mathbf{k}\cdot\mathbf{x}-\phi_{0}).\label{eq:V-classical}
\end{equation}
where $\hat{\boldsymbol{d}}=\vec{\wp}\,(\hat{\sigma}^{-}+\hat{\sigma}^{+})$
is the dipole moment operator of the TLS, with $\vec{\wp}:=\langle\mathsf{e}|\hat{\boldsymbol{d}}|\mathsf{g}\rangle$
as the transition dipole moment, and $\hat{\sigma}^{+}:=|\mathsf{e}\rangle\langle\mathsf{g}|=(\hat{\sigma}^{-})^{\dagger}$.

In such an interaction, the driving light is indeed modeled as a planar
wave as in classical physics. Thus, if the driving light carries different
photon statistics (e.g., Poisson, sub-Poisson, thermal \citep{gerry_introductory_2005,sperling_sub-binomial_2012,teuber_nonclassical_2015,li_photon_2020,agarwal_quantum_2012}),
the above quasi-classical driving interaction cannot reflect this
difference. 

Recently, it was noticed that the different types of the input photon
statistics do exhibit significant features when they interact with
the same quantum system. For example, the squeezed light (with sub-Poissonian
photon statistics) could enhance the two-photon absorption fluorescence
by $\sim47$ times comparing with the normal laser light with the
same intensity \citep{li_squeezed_2020}, and also can be used to
exceed the cooling limit in the laser cooling experiments \citep{clark_sideband_2017,schafermeier_quantum_2016,saxena_laser_2006},
and different nonclassical light states may lead to significant differences
in fluorescence spectrum \citep{vyas_resonance_1992} and electron
transport \citep{souquet_photon-assisted_2014}. Thus, nonclassical
light driving may also bring in potential enhancements in more different
physics problems. However, that requires a more precise quantum description
for the light-matter interaction beyond the above quasi-classical
driving, which has not yet been developed well enough.

In this paper, we make a quantum approach to study the interaction
between a quantum system and a driving light, by which the specific
photon statistics of the incoming light flux can be taken into account.
Based on the interaction between a TLS and the  quantized EM field,
if the driving mode starts from a coherent state $|\alpha\rangle$
as its initial state, it turns out the system dynamics can be described
by a master equation, which just returns the above quasi-classical
driving widely adopted in literature.

Further, if the initial state of the driving mode is not a coherent
state, but a generic quantum state represented by a \emph{P} function
$\hat{\varrho}=\int d^{2}\alpha\,P(\alpha)|\alpha\rangle\langle\alpha|$,
it turns out the system dynamics can be rewritten as the \emph{P}
function average of many evolution ``branches'': in each dynamics
branch the driving mode starts from a coherent state, thus again it
can be solved separately as the above quasi-classical driving situation,
and then their \emph{P} function average gives the full dynamics.

Based on this approach, we study a photoelectric converter model \citep{galperin_current-induced_2005,rutten_reaching_2009,berbezier_photovoltaic_2013,wang_optimal_2014,su_photoelectric_2016,aroutiounian_quantum_2001,scully_quantum_2010},
and calculate the photoelectric currents generated by the input light
with different photon statistics (Poisson, sub-Poisson, thermal).
We find that the photoelectric currents generated from different input
photon statistics do exhibit significant differences, even if they
have the same light intensity. We prove that, among all the \emph{classical
light states} (those who have non-singular positive \emph{P} functions
\citep{gerry_introductory_2005,agarwal_quantum_2012}), the input
light with Poisson statistics generates the largest photoelectric
current; on the other hand, the current generated from a nonclassical
light with sub-Poisson statistics is even larger than this classical
limit. 

The paper is arranged as follows. In section 2, we discuss how the
quasi-classical approach can be derived from a  quantum treatment
when the driving light is a coherent state. In section 3, we discuss
how to study the system dynamics when the driving light is a generic
state. In section 4, we consider a photoelectric converter model and
study the photoelectric current by the quasi-classical approach. In
section 5, we study the photoelectric current generated by different
light states. The summary is drawn in section 6.

\section{Quantum treatment of quasi-classical driving }

First we show how the above quasi-classical interaction (\ref{eq:V-classical})
can be derived from a quantum treatment. We start from the general
interaction between the TLS and the  quantized EM field ($\hat{H}_{\text{\textsc{b}}}=\sum_{\mathbf{k},\varsigma}\hbar\omega_{\mathbf{k}}\hat{a}_{\mathbf{k}\varsigma}^{\dagger}\hat{a}_{\mathbf{k}\varsigma}$),
which reads (in the interaction picture\footnote{Throughout the paper,
$\hat{o}$ denotes the operator in the Schr\"odinger picture, and
$\tilde{o}(t)$ indicates the interaction picture. }) 
\begin{eqnarray}
\tilde{H}_{\text{\textsc{sb}}} & = & -\tilde{\boldsymbol{d}}(t)\cdot\tilde{\mathbf{E}}(\mathbf{x},t)\nonumber \\
 & = & -\sum_{\mathbf{k}\varsigma}\tilde{\boldsymbol{d}}(t)\cdot\hat{\mathrm{e}}_{\mathbf{k}\varsigma}\sqrt{\frac{\hbar\omega_{\mathbf{k}}}{2\epsilon_{0}V}}\Big[i\hat{a}_{\mathbf{k}\varsigma}e^{i\mathbf{k}\cdot\mathbf{x}-i\omega_{\mathbf{k}}t}+\text{h.c.}\Big],\label{eq:H-SB}
\end{eqnarray}
where $\varsigma$ is the polarization index of the EM field, and
$\mathbf{x}$ is the position of the TLS.

The initial state of the EM field is set as follows: a specific $(\mathbf{k}_{0}\varsigma_{0})$-mode
(the driving mode) starts from a coherent state $|\alpha\rangle_{\mathbf{k}_{0}\varsigma_{0}}$
($\alpha\equiv|\alpha|e^{i\phi_{\alpha}}$), while all the other modes
start from the vacuum state, i.e.,\begin{equation}
\hat{\boldsymbol{\rho}}_{\mathnormal{\textsc{b}}}^{(\alpha)_{\mathbf{k}_{0}\varsigma_{0}}}(0)=\bigotimes_{\mathbf{k}\varsigma}\hat{\varrho}_{\mathbf{k}\varsigma},\qquad
\hat{\varrho}_{\mathbf{k}\varsigma}
=\cases{|\alpha\rangle\langle\alpha|, & $(\mathbf{k}_{0}\varsigma_{0})$\text{-mode} \\  |0\rangle\langle0|, & \text{other modes} \\} \label{eq:Rho-SB-0}
\end{equation}

Under this initial state, the field operator $\hat{a}_{\mathbf{k}\varsigma}$
can be divided as its displacement and the vacuum fluctuation $\hat{a}_{\mathbf{k}\varsigma}=\langle\hat{a}_{\mathbf{k}\varsigma}\rangle+\delta\hat{a}_{\mathbf{k}\varsigma}$,
namely, the driving mode gives $\hat{a}_{\mathbf{k}_{0}\varsigma_{0}}=\alpha+\delta\hat{a}_{\mathbf{k}_{0}\varsigma_{0}}$
and the other modes give $\hat{a}_{\mathbf{k}\varsigma}=\delta\hat{a}_{\mathbf{k}\varsigma}$.
Then the interaction (\ref{eq:H-SB}) can be rewritten as $\tilde{H}_{\text{\textsc{sb}}}=\tilde{V}_{\alpha}(t)+\tilde{H}_{\text{\textsc{sb}}}^{(0)}$,
where 
\begin{eqnarray}
\tilde{V}_{\alpha}(t) & = & -\tilde{\boldsymbol{d}}(t)\cdot\vec{E}_{\alpha}\sin(\omega_{\mathbf{k}_{0}}t-\mathbf{k}_{0}\cdot\mathbf{x}-\phi_{\alpha}),\label{eq:V-alpha}\\
\tilde{H}_{\text{\textsc{sb}}}^{(0)} & = & -\sum_{\mathbf{k}\varsigma}\tilde{\boldsymbol{d}}(t)\cdot\hat{\mathrm{e}}_{\mathbf{k}\varsigma}\sqrt{\frac{\hbar\omega_{\mathbf{k}}}{2\epsilon_{0}V}}\Big[i\,\delta\hat{a}_{\mathbf{k}\varsigma}e^{i\mathbf{k}\cdot\mathbf{x}-i\omega_{\mathbf{k}}t}+\text{h.c.}\Big],\nonumber 
\end{eqnarray}
with $\vec{E}_{\alpha}:=\hat{\mathrm{e}}_{\mathbf{k}_{0}\varsigma_{0}}|\alpha|\sqrt{2\hbar\omega_{\mathbf{k}_{0}}/\epsilon_{0}V}$
(set $\mathbf{x}\equiv0$ hereafter). 

Therefore, $\tilde{V}_{\alpha}(t)$ just gives the above quasi-classical
interaction (\ref{eq:V-classical}) between the TLS and a planar wave.
We remark that up to now the above treatments are exact without any
rotating-wave approximation (RWA), and it applies for both resonant
and non-resonant driving.

On the other hand, in the interaction term $\tilde{H}_{\text{\textsc{sb}}}^{(0)}$
of equation (\ref{eq:V-alpha}), $\delta\hat{a}_{\mathbf{k}\varsigma}=\hat{a}_{\mathbf{k}\varsigma}-\langle\hat{a}_{\mathbf{k}\varsigma}\rangle$
only contains the field fluctuation around its mean value, which satisfies
$\langle\delta\hat{a}_{\mathbf{k}\varsigma}\rangle=0$, $\langle\delta\hat{a}_{\mathbf{k}\varsigma}\,\delta\hat{a}_{\mathbf{k}'\varsigma'}^{\dagger}\rangle=\delta_{\mathbf{k}\mathbf{k}'}\delta_{\varsigma\varsigma'}$,
and $\langle\delta\hat{a}_{\mathbf{k}\varsigma}^{\dagger}\delta\hat{a}_{\mathbf{k}'\varsigma'}^{\dagger}\rangle=\langle\delta\hat{a}_{\mathbf{k}\varsigma}\,\delta\hat{a}_{\mathbf{k}'\varsigma'}\rangle=0$
for all ($\mathbf{k}\varsigma$)-modes. Notice that these relations
and $\tilde{H}_{\text{\textsc{sb}}}^{(0)}$ just have the same form
as the weak interaction between the TLS and the quantized vacuum field
when considering the spontaneous emission. Thus we can apply the Born-Markovian
approximation and RWA \citep{breuer_theory_2002}, and obtain the
following master equation of the system dynamics (see derivation in
\ref{sec:Master-equation-derivation}) 
\begin{eqnarray}
\partial_{t}\tilde{\rho}_{\text{\textsc{s}}}^{(\alpha)} & = & \frac{i}{\hbar}[\tilde{\rho}_{\text{\textsc{s}}}^{(\alpha)},\,\tilde{V}_{\alpha}(t)]+{\cal L}_{\text{\textsc{em}}}[\tilde{\rho}_{\text{\textsc{s}}}^{(\alpha)}],\nonumber \\
{\cal L}_{\text{\textsc{em}}}[\rho] & = & \kappa\big(\hat{\sigma}^{-}\rho\hat{\sigma}^{+}-\frac{1}{2}\{\hat{\sigma}^{+}\hat{\sigma}^{-},\,\rho\}\big).\label{eq:ME}
\end{eqnarray}
This is just the master equation widely adopted in literature, which
contains both the quasi-classical driving and the spontaneous emission
term ${\cal L}_{\text{\textsc{em}}}[\rho]$ with decay rate $\kappa$.
But now the driving term here is no longer directly imposed from the
quasi-classical interaction (\ref{eq:V-classical}) in priori, but
emerges from the initial coherent state of the quantized field (\ref{eq:Rho-SB-0}).

\section{Driving by generic light states }

Now we consider a more general situation that the initial state of
the driving mode is not a coherent state but a generic quantum state,
while all the other modes still start from the vacuum state. 

In this case, such an initial state cannot return the above quasi-classical
driving any more. Generally, the initial states of the $(\mathbf{k}_{0}\varsigma_{0})$-mode
and the whole EM field can be written in the following \emph{P} representation
\citep{gerry_introductory_2005,agarwal_quantum_2012,sudarshan_equivalence_1963,glauber_coherent_1963,scully_quantum_1997,souquet_photon-assisted_2014,marte_lasers_1989},
\begin{eqnarray}
\hat{\varrho}_{\mathbf{k}_{0}\varsigma_{0}} & = & \int d^{2}\alpha\,P(\alpha)\,|\alpha\rangle_{\mathbf{k}_{0}\varsigma_{0}}\langle\alpha|,\nonumber \\
\hat{\boldsymbol{\rho}}_{\mathnormal{\textsc{b}}}(0) & = & \bigotimes_{\mathbf{k}\varsigma}\,\hat{\varrho}_{\mathbf{k}\varsigma}=\int d^{2}\alpha\,P(\alpha)\,\hat{\boldsymbol{\rho}}_{\mathnormal{\textsc{b}}}^{(\alpha)_{\mathbf{k}_{0}\varsigma_{0}}},\label{eq:General-Rho-B}
\end{eqnarray}
 where $\hat{\boldsymbol{\rho}}_{\mathnormal{\textsc{b}}}^{(\alpha)_{\mathbf{k}_{0}\varsigma_{0}}}$
is just given by equation\,(\ref{eq:Rho-SB-0}). The bath state $\hat{\boldsymbol{\rho}}_{\mathnormal{\textsc{b}}}(0)$
``looks like'' a probabilistic collection of many components $\hat{\boldsymbol{\rho}}_{\mathnormal{\textsc{b}}}^{(\alpha)_{\mathbf{k}_{0}\varsigma_{0}}}$,
but remember the \emph{P} function $P(\alpha)$ is not a probability
distribution and it may contain negative parts \citep{agarwal_quantum_2012,gerry_introductory_2005,scully_quantum_1997}.

Now the evolution of the system state $\hat{\rho}_{\text{\textsc{s}}}(t)$
can be given by 
\begin{eqnarray}
\hat{\rho}_{\mathnormal{\textsc{s}}}(t) & = & \mathrm{tr}_{\text{\textsc{b}}}\Big\{\mathcal{E}_{t}[\hat{\rho}_{\mathnormal{\textsc{s}}}(0)\otimes\hat{\boldsymbol{\rho}}_{\mathnormal{\textsc{b}}}(0)]\Big\}\nonumber \\
 & = & \int d^{2}\alpha\,P(\alpha)\,\mathrm{tr}_{\text{\textsc{b}}}\Big\{\mathcal{E}_{t}[\hat{\rho}_{\mathnormal{\textsc{s}}}(0)\otimes\hat{\boldsymbol{\rho}}_{\mathnormal{\textsc{b}}}^{(\alpha)}(0)]\Big\}\nonumber \\
 & := & \int d^{2}\alpha\,P(\alpha)\,\hat{\rho}_{\text{\textsc{s}}}^{(\alpha)}(t),\label{eq:rho-branch}
\end{eqnarray}
 where $\mathcal{E}_{t}[...]$ is the unitary evolution operator of
the whole \noun{s-b} system, and $\hat{\rho}_{\text{\textsc{s}}}^{(\alpha)}(t):=\mathrm{tr}_{\text{\textsc{b}}}\Big\{\mathcal{E}_{t}[\hat{\rho}_{\mathnormal{\textsc{s}}}(0)\otimes\hat{\boldsymbol{\rho}}_{\mathnormal{\textsc{b}}}^{(\alpha)_{\mathbf{k}_{0}\varsigma_{0}}}(0)]\Big\}$. 

It is worth noting that indeed $\hat{\rho}_{\text{\textsc{s}}}^{(\alpha)}(t)$
indicates the system dynamics when the field state starts from $\hat{\boldsymbol{\rho}}_{\mathnormal{\textsc{b}}}^{(\alpha)_{\mathbf{k}_{0}\varsigma_{0}}}(0)$
{[}equation\,(\ref{eq:Rho-SB-0}){]}, which is just the above situation
of quasi-classical driving given $|\alpha\rangle$ as the initial
state of the $(\mathbf{k}_{0}\varsigma_{0})$-mode. 

Therefore, the complete system dynamics $\hat{\rho}_{\mathnormal{\textsc{s}}}(t)$
{[}equation (\ref{eq:rho-branch}){]} can be regarded as the \emph{P}
function average of many evolution ``branches'' $\hat{\rho}_{\text{\textsc{s}}}^{(\alpha)}(t)$,
and we call $\hat{\rho}_{\text{\textsc{s}}}^{(\alpha)}(t)$ as the
$\alpha$\emph{-branch} of the full dynamics. In each $\alpha$-branch,
the driving mode just starts from the coherent state $|\alpha\rangle$
as the initial state, thus, the system dynamics can be regarded as
governed by the quasi-classical driving interaction $\tilde{V}_{\alpha}(t)$
and the weak coupling with the quantized vacuum field $\tilde{H}_{\text{\textsc{sb}}}^{(0)}$
{[}equation (\ref{eq:V-alpha}){]}. Approximately, $\hat{\rho}_{\mathnormal{\textsc{s}}}(t)$
can be given by the above master equation (\ref{eq:ME}) from Born-Markovian
approximation and RWA.

Besides, equation (\ref{eq:rho-branch}) also provides a simple way
to obtain the dynamics of system observable expectations, i.e.,
\begin{equation}
\langle\hat{O}_{\mathnormal{\textsc{s}}}(t)\rangle=\mathrm{tr}_{\text{\textsc{s}}}\big[\hat{O}_{\mathnormal{\textsc{s}}}\,\hat{\rho}_{\text{\textsc{s}}}(t)\big]=\int d^{2}\alpha\,P(\alpha)\,\langle\hat{O}_{\mathnormal{\textsc{s}}}(t)\rangle^{(\alpha)},\label{eq:Os}
\end{equation}
where $\langle\hat{O}_{\mathnormal{\textsc{s}}}(t)\rangle^{(\alpha)}=\mathrm{tr}_{\text{\textsc{s}}}[\hat{O}_{\mathnormal{\textsc{s}}}\cdot\hat{\rho}_{\text{\textsc{s}}}^{(\alpha)}(t)]$
can be obtained by the master equation (\ref{eq:ME}) with quasi-classical
driving.

In sum, if the driving light on the system is not a coherent state,
the system dynamics $\langle\hat{O}_{\mathnormal{\textsc{s}}}(t)\rangle$
can be obtained as the \emph{P} function average of all the branches
$\langle\hat{O}_{\mathnormal{\textsc{s}}}(t)\rangle^{(\alpha)}$,
and each branch can be given by the master equation (\ref{eq:ME})
with the quasi-classical driving interaction. 

We emphasize that, the interaction (\ref{eq:V-alpha}) in each $\alpha$-branch
and the \emph{P} function averages (\ref{eq:rho-branch}, \ref{eq:Os})
are formally exact, but evaluating the dynamics $\hat{\rho}_{\text{\textsc{s}}}^{(\alpha)}(t)$
of each $\alpha$-branch often requires some approximations (e.g.,
Born-Markovian approximation, and RWA). When the system-bath coupling
strength is strong, the backaction from the system to the field could
be important; in this case, high-order Markovian corrections can be
taken into account in the evaluation of each $\alpha$-branch, and
the above \emph{P} function averages (\ref{eq:rho-branch}, \ref{eq:Os})
still apply. Throughout this paper, we focus on the situation that
the system-bath coupling is quite weak, and quantum effect only comes
from the input light state, thus the above Markovian master equation
is precise enough for each $\alpha$-branch. 

If more than one TLS are concerned, the generalization is straightforward:
in each $\alpha$-branch, the interaction between each single TLS
with the EM field is still given by the interaction (\ref{eq:V-alpha}),
which can be further used to study the field induced interaction between
different TLSs \citep{lehmberg_radiation_1970,agarwal_quantum_1974,ficek_entangled_2002,wang_magnetic_2018,hu_field-induced_2020}.
Throughout this paper we only focus on the situation of one TLS, and
do not consider the field induced interaction.

\section{Photoelectric converter model}

\begin{figure}
\centering{}\includegraphics[width=0.4\columnwidth]{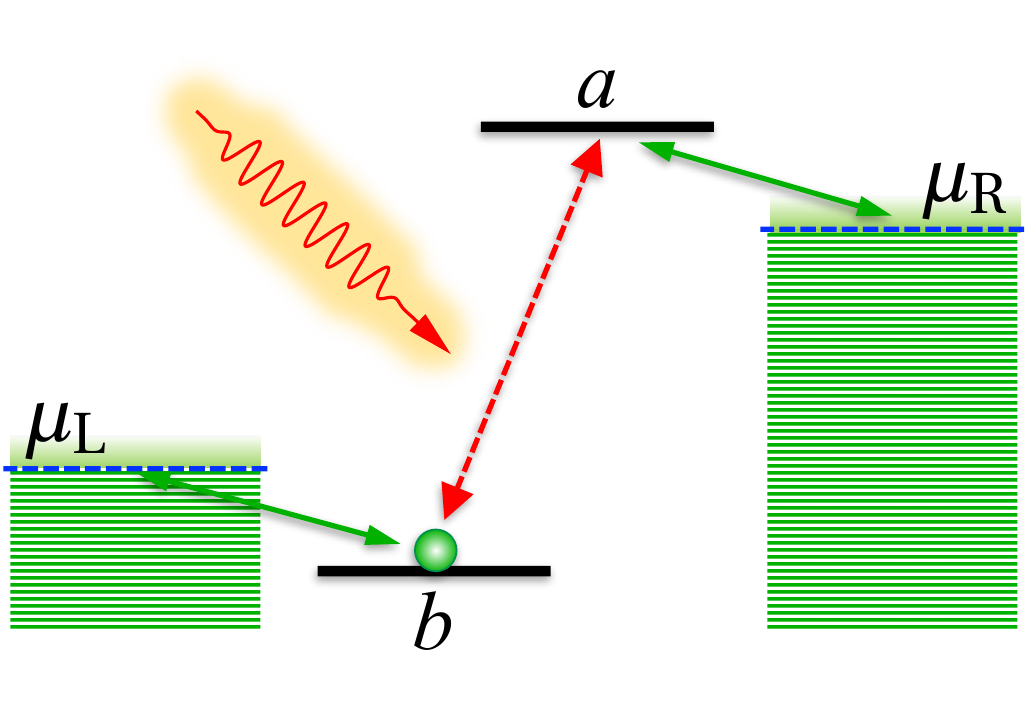}\caption{\label{fig-QD}Demonstration of the photoelectric converter model.
The fermionic level-$a(b)$ is coupled to the right (left) electron
lead, whose chemical potential is $\mu_{\textsc{r}}$ ($\mu_{\text{\textsc{l}}}$),
and $\hbar\Omega_{a}>\mu_{\text{\textsc{r}}}>\mu_{\text{\textsc{l}}}>\hbar\Omega_{b}\equiv0$.
The incoming photons excite the electron across the voltage barrier
and generate the photoelectric current.}
\end{figure}

Now we consider a photoelectric converter model  and study the photoelectric
current excited from different light states. The photoelectric converter
is modeled as two fermionic levels, $\hat{H}_{\mathnormal{\textsc{s}}}=\hbar\Omega_{a}\hat{a}^{\dagger}\hat{a}+\hbar\Omega_{b}\hat{b}^{\dagger}\hat{b}$
(setting $\Omega_{b}\equiv0$, and $\Omega_{a}-\Omega_{b}:=\Omega$),
and they contact with two electron leads $\hat{H}_{\text{\textsc{l}}(\text{\textsc{r}})}=\sum_{k}\varepsilon_{\text{\textsc{l}}(\text{\textsc{r}}),k}\,\hat{c}_{\text{\textsc{l}}(\text{\textsc{r}}),k}^{\dagger}\hat{c}_{\text{\textsc{l}}(\text{\textsc{r}}),k}$
respectively via the tunneling interaction $\hat{V}_{\text{\textsc{l}}}=\sum_{k}g_{\text{\textsc{l}},k}\,\hat{b}^{\dagger}\hat{c}_{\text{\textsc{l}},k}+\text{h.c.}$
and $\hat{V}_{\text{\textsc{r}}}=\sum_{k}g_{\text{\textsc{r}},k}\,\hat{a}^{\dagger}\hat{c}_{\text{\textsc{r}},k}+\text{h.c.}$
(see figure \ref{fig-QD}, here $\hat{a}$, $\hat{b}$, $\hat{c}_{\text{\textsc{l}}(\text{\textsc{r}}),k}$
are the fermionic annihilation operators of the two levels and the
electron modes in the leads). This model has been well used to study
photoelectric current generation in a solar cell \citep{rutten_reaching_2009,berbezier_photovoltaic_2013,wang_optimal_2014,su_photoelectric_2016,aroutiounian_quantum_2001,childress_mesoscopic_2004},
and the photon-induced electron transport across a molecule junction
\citep{galperin_current-induced_2005,wu_atomic-scale_2006,zhou_photoconductance_2018}.

In a photoelectric converter made by a p-n diode, the different doping
types make the region near the p-n\emph{ }interface lose the electric
neutrality and form a depletion layer, and that creates an internal
electric field, which makes the diode unidirectional \citep{aroutiounian_quantum_2001}.
Thus here the two fermionic levels do not have direct tunneling, and
they cannot exchange with each other without the mediation of the
EM field. The incoming photons could stimulate the electron up and
down, exchanging between these two levels. The interaction between
these two fermion levels and the quantized EM field is just the above
$\hat{H}_{\text{\textsc{sb}}}$ {[}equation (\ref{eq:H-SB}){]}, except
here the dipole moment operator should be modified as $\hat{\boldsymbol{d}}=\vec{\wp}(\hat{\tau}^{-}+\hat{\tau}^{+})$
with $\hat{\tau}^{+}:=\hat{a}^{\dagger}\hat{b}=(\hat{\tau}^{-})^{\dagger}$.

Therefore, the above discussions for different input light states
can be well applied here. We first consider the situation that the
driving light is a coherent state $|\alpha\rangle$ {[}equation (\ref{eq:Rho-SB-0}){]},
then the master equation for the system dynamics is obtained as 
\begin{eqnarray}
\partial_{t}\tilde{\rho}_{\text{\textsc{s}}} & = & \frac{i}{\hbar}[\tilde{\rho}_{\text{\textsc{s}}},\tilde{V}_{\alpha}(t)]+{\cal L}_{\text{\textsc{em}}}[\tilde{\rho}_{\text{\textsc{s}}}]+{\cal L}_{a}[\tilde{\rho}_{\text{\textsc{s}}}]+{\cal L}_{b}[\tilde{\rho}_{\text{\textsc{s}}}],\label{eq:QD-ME}\\
\tilde{V}_{\alpha}(t) & = & i\hbar\xi_{0}\alpha\,\hat{\tau}^{+}e^{i(\Omega-\omega_{\mathbf{k}_{0}})t}-i\hbar\xi_{0}^{*}\alpha^{*}\,\hat{\tau}^{-}e^{-i(\Omega-\omega_{\mathbf{k}_{0}})t}.\nonumber 
\end{eqnarray}
 Here RWA has been applied to the driving interaction $\tilde{V}_{\alpha}(t)$,
and $\hbar\xi_{0}:=-(\vec{\wp}\cdot\hat{\mathrm{e}}_{\mathbf{k}_{0}\varsigma_{0}})\sqrt{\hbar\omega_{\mathbf{k}_{0}}/2\epsilon_{0}V}$
is the single-photon coupling strength. Hereafter we only focus on
the resonant driving case and set $\omega_{\mathbf{k}_{0}}\equiv\Omega$.

${\cal L}_{\text{\textsc{em}}}[\tilde{\rho}_{\text{\textsc{s}}}]$
is the same with equation (\ref{eq:ME}) except here $\hat{\sigma}^{\pm}$
should be replaced by $\hat{\tau}^{\pm}$, which describes the spontaneous
emission. ${\cal L}_{a(b)}[\tilde{\rho}_{\text{\textsc{s}}}]$ describes
the dissipation due to coupling with the right (left) lead, which
reads (taking $\text{\textsc{q}}=a,b$) \citep{su_photoelectric_2016,wiseman_quantum_2010,galperin_current-induced_2005,schaller_open_2014,wang_optimal_2014}
\begin{eqnarray}
\mathcal{L}_{\text{\textsc{q}}}[\rho] & = & \gamma_{\text{\textsc{q}}}\bar{\mathtt{n}}_{\text{\textsc{q}}}(\hat{\text{\textsc{q}}}^{\dagger}\rho\hat{\text{\textsc{q}}}-\frac{1}{2}\hat{\text{\textsc{q}}}\hat{\text{\textsc{q}}}^{\dagger}\rho-\frac{1}{2}\rho\hat{\text{\textsc{q}}}\hat{\text{\textsc{q}}}^{\dagger})\nonumber \\
 & + & \gamma_{\text{\textsc{q}}}(1-\bar{\mathtt{n}}_{\text{\textsc{q}}})(\hat{\text{\textsc{q}}}\rho\hat{\text{\textsc{q}}}^{\dagger}-\frac{1}{2}\hat{\text{\textsc{q}}}^{\dagger}\hat{\text{\textsc{q}}}\rho-\frac{1}{2}\rho\hat{\text{\textsc{q}}}^{\dagger}\hat{\text{\textsc{q}}}),
\end{eqnarray}
where $\bar{\mathtt{n}}_{a(b)}=\big[\exp\beta_{\text{\textsc{r(l)}}}(\hbar\Omega_{a(b)}-\mu_{\text{\textsc{r(l)}}})+1\big]^{-1}$
is the Fermi-Dirac distribution, and $\mu_{\text{\textsc{r(l)}}}$
is the chemical potential of the right (left) lead. Here we consider
the temperatures of the two electron leads are zero, which gives $\bar{\mathtt{n}}_{b}=1$,
$\bar{\mathtt{n}}_{a}=0$.

From the master equation (\ref{eq:QD-ME}), the average electron number
$\langle\hat{\text{\textsc{n}}}_{a}\rangle:=\langle\hat{a}^{\dagger}\hat{a}\rangle$
on level-$a$ gives 
\begin{eqnarray}
\partial_{t}\langle\hat{\text{\textsc{n}}}_{a}\rangle & = & \mathrm{tr}\Big\{\frac{i}{\hbar}[\tilde{\rho}_{\mathnormal{\textsc{s}}},\tilde{V}_{\alpha}]\,\hat{\text{\textsc{n}}}_{a}+\mathcal{L}_{\mathnormal{\textsc{em}}}[\tilde{\rho}_{\mathnormal{\textsc{s}}}]\,\hat{\text{\textsc{n}}}_{a}\Big\}+\mathrm{tr}\big\{\mathcal{L}_{a}[\tilde{\rho}_{\mathnormal{\textsc{s}}}]\,\hat{\text{\textsc{n}}}_{a}\big\}\nonumber \\
 & := & J_{\text{\textsc{em}}}-J_{\text{\textsc{r}}}.
\end{eqnarray}
Here $J_{\text{\textsc{r}}}:=-\mathrm{tr}\big\{\mathcal{L}_{a}[\tilde{\rho}_{\mathnormal{\textsc{s}}}]\,\hat{\text{\textsc{n}}}_{a}\big\}$
is the current flowing from level-$a$ to the right lead, and $J_{\text{\textsc{em}}}$
is the net exciting rate from level-$b$ to level-$a$. In the steady
state $\partial_{t}\langle\hat{\text{\textsc{n}}}_{a}\rangle\big|_{t\rightarrow\infty}=0$,
we have $J_{\text{\textsc{em}}}=J_{\text{\textsc{r}}}:=J(\alpha)$,
and the photoelectric current is $-e\,J_{\text{\textsc{r}}}$.

The spontaneous rate is usually much smaller than the tunneling rates
$\kappa\ll\gamma_{a,b}:=\gamma$. The above steady state current can
be obtained from the master equation (\ref{eq:QD-ME}) {[}see equation
(\ref{eq:J-R}) in \ref{sec:Photoelectric-current}{]}
\begin{equation}
J(\alpha)=\frac{2|\xi_{0}|^{2}\,|\alpha|^{2}\,\gamma}{4|\xi_{0}|^{2}\,|\alpha|^{2}+\gamma^{2}}=\frac{\gamma}{2}\Big[1-\frac{\tilde{\gamma}_{\xi}^{2}}{4|\alpha|^{2}+\tilde{\gamma}_{\xi}^{2}}\Big],\label{eq:J-alpha}
\end{equation}
where $\tilde{\gamma}_{\xi}:=\gamma/|\xi_{0}|$. Thus a non-zero input
light ($\alpha\neq0$) always produces a photoelectric current across
the voltage barrier {[}$J(\alpha)>0$ means the electrons move from
left to right{]}.

\section{Photoelectric current generated by different light states}

Now we consider the driving light is not a coherent state, which is
beyond the previous quasi-classical description. In this case, equation\,(\ref{eq:J-alpha})
just gives the steady current for the $\alpha$-branch dynamics, and
the complete result should be the summation from all branches {[}equation\,(\ref{eq:Os}){]},
that is, $\overline{J}:=\int d^{2}\alpha\,P(\alpha)J(\alpha)$.

When the light intensity is weak ($|\alpha|^{2}\ll\tilde{\gamma}_{\xi}^{2}\equiv\gamma^{2}/|\xi_{0}|^{2}$),
the current equation (\ref{eq:J-alpha}) gives $J(\alpha)\simeq(2|\xi_{0}|^{2}/\gamma)\,|\alpha|^{2}$,
thus its \emph{P} function average always gives the full steady current
as $\overline{J}=(2|\xi_{0}|^{2}/\gamma)\,\overline{n}$. That means,
the photoelectric current is always proportional to the average photon
number $\overline{n}$ (namely, the light intensity) in spite of the
input photon statistics. If this weak intensity condition is not satisfied,
the photoelectric current may exhibit significant differences for
different input light states.

\begin{figure}
\centering{}\includegraphics[width=0.8\columnwidth]{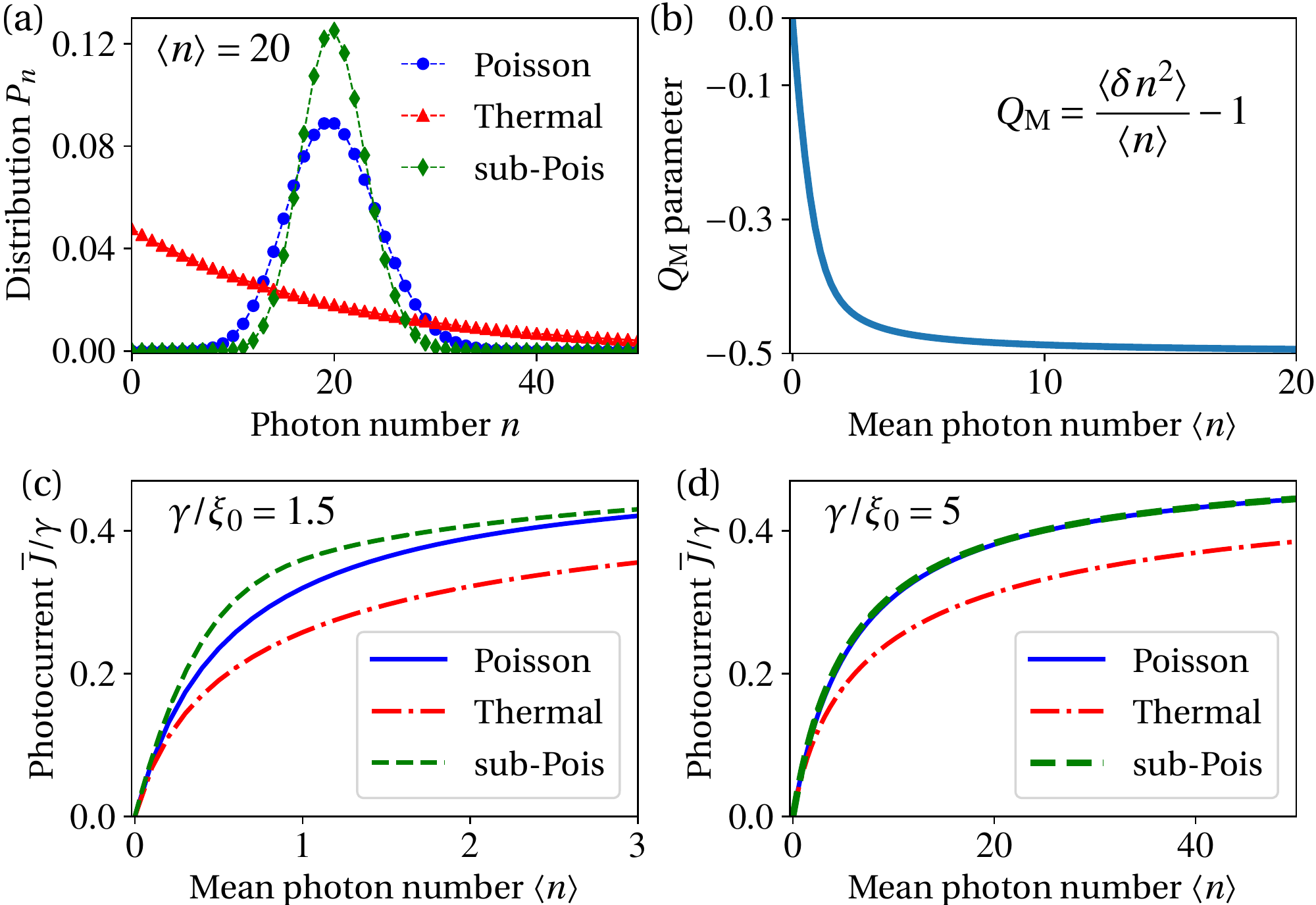}\caption{(a) Photon number distribution $P_{n}$ for the thermal, Poisson,
sub-Poisson {[}equation (\ref{eq:Pn-subP}){]} statistics with the
same mean photon number $\overline{n}=20$. (b) The Mandel $Q_{\text{M}}$
parameter for the sub-Poisson distribution {[}equation (\ref{eq:Pn-subP}){]}
under different mean photon number. (c, d) The photoelectric current
$\overline{J}/\gamma$ generated by the Poisson, thermal, sub-Poisson
light {[}equations (\ref{eq:J-alpha}, \ref{eq:J-th}, \ref{eq:J_sub}){]}
(given $\tilde{\gamma}_{\xi}^{2}\equiv\gamma/|\xi_{0}|=1.5,\,5$).}
\label{fig-J}
\end{figure}

We first consider the input light state is a uniform mixture of all
the coherent state $|\alpha\rangle$ with the same photon number $|\alpha|^{2}\equiv\overline{n}$
but different phases $\phi_{\alpha}$, which can be written as $\rho=\int\frac{d\phi_{\alpha}}{2\pi}\,|\alpha\rangle\langle\alpha|=\sum P_{n}|n\rangle\langle n|$,
with $P_{n}=e^{-|\alpha|^{2}}|\alpha|^{2n}/n!$ as the Poisson distribution.
In this situation (the idealistic laser statistics), the \emph{P}
function average on $J(\alpha)$ gives the same result as equation
(\ref{eq:J-alpha}) {[}solid blue line in figure \ref{fig-J}(c, d){]}.

Now we consider the input light is a monochromatic one carrying the
thermal statistics, described by the \emph{P }function $P_{\text{th}}(\alpha)=[\pi\bar{n}]^{-1}\exp[-|\alpha|^{2}/\bar{n}]$
with $\overline{n}$ as the mean photon number \citep{agarwal_quantum_2012,li_photon_2020,gerry_introductory_2005,scully_quantum_1997}.
In this case, the steady current becomes 
\begin{equation}
\overline{J}_{\text{th}}=\int d^{2}\alpha\,P_{\text{th}}(\alpha)J(\alpha)=\frac{\gamma}{2}\Big[1+\frac{\tilde{\gamma}_{\xi}^{2}}{4\overline{n}}\,e^{\frac{\tilde{\gamma}_{\xi}^{2}}{4\overline{n}}}\mathrm{Ei}(-\frac{\tilde{\gamma}_{\xi}^{2}}{4\overline{n}})\Big],\label{eq:J-th}
\end{equation}
 where $\mathrm{Ei}(x):=-\int_{-x}^{\infty}dt\,e^{-t}/t$ is the exponential
integral function {[}chain red line in figure \ref{fig-J}(c, d){]}. 

It turns out that, under the same average photon number (light intensity),
the currents excited from the Poisson and thermal light exhibit significant
differences. The current generated by the Poisson light is always
larger than the thermal case {[}figure \ref{fig-J}(c, d){]}. Meanwhile,
in the weak intensity region ($0<\overline{n}\ll\tilde{\gamma}_{\xi}^{2}$),
these two results {[}equations (\ref{eq:J-alpha}, \ref{eq:J-th}){]}
almost coincide with each other, and both exhibit a linear dependence
on the average photon number $\overline{n}$, which is consistent
with the above discussions.

Further, with the help of Lagrangian multipliers, we can prove, among
all the classical light states (those who have $P(\alpha)\ge0$),
under the same mean photon number $\overline{n}$, the Poisson input
generates the largest photoelectric current $\overline{J}=\int d^{2}\alpha\,P(\alpha)J(\alpha)$
(\ref{sec:Proof-for-the}). Namely, the photoelectric current generated
from the Poisson light sets the upper bound for all classical light
states.

Now we consider the driving light has the following sub-Poisson statistics,
\begin{eqnarray}
P_{n} & = & \frac{1}{I_{0}(2\sqrt{\lambda})}\frac{\lambda^{n}}{(n!)^{2}},\nonumber \\
\overline{n} & = & \frac{\sqrt{\lambda}\,I_{1}(2\sqrt{\lambda})}{I_{0}(2\sqrt{\lambda})},\qquad\overline{n^{2}}=\lambda,\label{eq:Pn-subP}
\end{eqnarray}
 where $I_{0/1}(x)$ is the modified Bessel function of the first
kind.

The distribution profile is shown in figure \ref{fig-J}(a) (green
diamonds, for $\overline{n}=20$), and clearly it is narrower than
the Poisson distribution with the same average photon number (blue
dots). The Mandel $Q_{\text{M}}$\emph{ }parameter ($Q_{\text{M}}:=\langle\delta n^{2}\rangle/\langle n\rangle-1$)
of this distribution is always negative {[}figure \ref{fig-J}(b){]},
which means such a photon statistics is a nonclassical one, and its
\emph{P} function is not positive-definite \citep{gerry_introductory_2005,agarwal_quantum_2012,mandel_sub-poissonian_1979}. 

The photoelectric current generated by this sub-Poisson light can
be obtained by the \emph{P} function average of equation (\ref{eq:J-alpha}).
Notice that, this \emph{P} function average is also equivalent with
the normal-order expectation on the light state $\rho=\sum P_{n}|n\rangle\langle n|$
\citep{sudarshan_equivalence_1963,glauber_coherent_1963,scully_quantum_1997,gerry_introductory_2005,agarwal_quantum_2012},
namely, $\overline{J}=\langle:J\big(\alpha^{*}\rightarrow\hat{a}^{\dagger},\,\alpha\rightarrow\hat{a}\big):\rangle$,
where $\langle:J(\hat{a}^{\dagger},\hat{a}):\rangle$ means the normal-order
expectation. This can be further calculated with the help of Widder
transform \citep{agarwal_quantum_2012,widder_convolution_1954} (\ref{sec:Generic-input-photon}),
which gives the steady state current as 
\begin{equation}
\overline{J}_{\text{sub}}=\frac{\gamma}{2}\Big[1-\tilde{\gamma}_{\xi}^{2}\int_{0}^{\infty}ds\,e^{-\tilde{\gamma}_{\xi}^{2}s}\,\frac{I_{0}(2\sqrt{(1-4s)\lambda})}{I_{0}(2\sqrt{\lambda})}\Big].\label{eq:J_sub}
\end{equation}

The photoelectric current generated by such a sub-Poisson light is
shown in figure \ref{fig-J}(c, d) (dashed green line), and it is
larger than the above classical upper bound set by the Poisson light.
Notice that the surpassing amount is dependent on the tunneling rate
$\gamma$ comparing with the single-photon coupling strength $\xi_{0}$.
In most practical situations $\gamma\gg|\xi_{0}|$, this difference
is quite small {[}figure \ref{fig-J}(d){]}. If the tunneling rate
is small ($\gamma\sim\xi_{0}$), such a difference due to the input
photon statistics could be significant. On the other hand, the difference
between the currents generated by the thermal and Poisson light appears
independent on $\gamma/|\xi_{0}|\equiv\tilde{\gamma}_{\xi}$ {[}indeed
in both equations (\ref{eq:J-alpha}, \ref{eq:J-th}), $\overline{n}/\tilde{\gamma}_{\xi}^{2}$
appears together as a whole{]}. 

It is known that the Poissonian distribution indicates the photons
are arriving randomly, while the sub-Poisson light exhibits the anti-bunching
effect, indicating the photons are arriving more ``regularly'' than
completely random \citep{gerry_introductory_2005,scully_quantum_1997,agarwal_quantum_2012},
which leads to the above enhancement. Clearly, nonclassical states
are a much larger set than the classical ones, and anti-bunching is
just one particular kind of quantum features, thus it is possible
that different kinds of nonclassical light may lead to some other
novel effects.

\section{Summary }

In this paper, we made a quantum approach to study photoelectric current
generated by a monochromatic driving light which carries a generic
photon statistics. If the driving mode starts from a coherent state
as the initial state, our quantum treatment just returns the quasi-classical
driving description as widely adopted in literature. But if the driving
light has a generic photon statistics with a given \emph{P} function,
the full system dynamics becomes the \emph{P} function average of
many evolution ``branches'': in each dynamics branch, the driving
mode starts from a coherent state and thus returns the quasi-classical
driving. Based on this quantum approach, it turns out, different types
of photon statistics do make differences to the photoelectric current
generation. Among all the classical light states with the same mean
photon number, the Poisson statistics generates the largest photoelectric
current, while a nonclassical sub-Poisson light could even exceed
this classical upper bound. The sub-Poissonian driving light may be
realized by the squeezed light or sub-Poissonian laser \citep{golubev_photon_1984,richardson_nonclassical_1990,davidovich_sub-poissonian_1996,wiersig_direct_2009}.
The model here has been used to study the photon-induced electron
transport in quantum dots \citep{berbezier_photovoltaic_2013,liu_threshold_2017}
and molecule junctions \citep{galperin_current-induced_2005,wu_atomic-scale_2006,zhou_photoconductance_2018}.
In principle the above novel results in our study could be observable
in these platform when the tunneling rate $\gamma$ is small enough.
Meanwhile, it is expectable that some other quantum states which may
lead to stronger enhancement in such electronic transport systems,
and this approach also can be applied in more different problems with
light driving.

\ack S.-W. Li appreciates quite much for the helpful discussion with
Y. Li in CSRC. This study is supported by NSF of China (Grant No.11905007),
Beijing Institute of Technology Research Fund Program for Young Scholars.

\appendix

\section{Master equation derivation \label{sec:Master-equation-derivation}}

Here we present the derivation for the the master equation (\ref{eq:ME})
in the main text. Since the EM field starts from $\hat{\boldsymbol{\rho}}_{\mathnormal{\textsc{b}}}^{(\alpha)_{\mathbf{k}_{0}\varsigma_{0}}}(0)$
{[}equation (\ref{eq:Rho-SB-0}) in the main text{]}, in the interaction
picture, the interaction between the two-level system and the EM field
can be rewritten as $\tilde{H}_{\text{\textsc{sb}}}=\tilde{V}_{\alpha}(t)+\tilde{H}_{\text{\textsc{sb}}}^{(0)}$
{[}equation (\ref{eq:V-alpha}) in the main text{]}, where $\tilde{V}_{\alpha}(t)=-\tilde{\boldsymbol{d}}(t)\cdot\vec{E}_{\alpha}\sin(\omega_{\mathbf{k}_{0}}t-\mathbf{k}_{0}\cdot\mathbf{x}-\phi_{\alpha})$,
and 
\begin{eqnarray}
\tilde{H}_{\text{\textsc{sb}}}^{(0)} & =-\sum_{\mathbf{k},\varsigma} & \big(\hat{\sigma}^{-}e^{-i\Omega t}+\hat{\sigma}^{+}e^{i\Omega t}\big)\cdot\nonumber \\
 &  & (\vec{\wp}\cdot\hat{\mathrm{e}}_{\mathbf{k}\varsigma})\sqrt{\frac{\hbar\omega_{\mathbf{k}}}{2\epsilon_{0}V}}\Big[i\,\delta\hat{a}_{\mathbf{k}\varsigma}\,e^{i\mathbf{k}\cdot\mathbf{x}-i\omega_{\mathbf{k}}t}+\text{h.c.}\Big].
\end{eqnarray}
 Here $\hat{\sigma}^{+}=|\mathsf{e}\rangle\langle\mathsf{g}|=(\hat{\sigma}^{-})^{\dagger}$,
and $\tilde{\boldsymbol{d}}(t)=\vec{\wp}(\hat{\sigma}^{-}e^{-i\Omega t}+\text{h.c.})$.
The operator $\delta\hat{a}_{\mathbf{k}\varsigma}=\hat{a}_{\mathbf{k}\varsigma}-\langle\hat{a}_{\mathbf{k}\varsigma}\rangle$
indicates the pure fluctuation of the quantized field, and the displacement
$\langle\hat{a}_{\mathbf{k}\varsigma}\rangle=\mathrm{tr}_{\text{\textsc{b}}}[\hat{\boldsymbol{\rho}}_{\mathnormal{\textsc{b}}}^{(\alpha)_{\mathbf{k}_{0}\varsigma_{0}}}(0)\,\hat{a}_{\mathbf{k}\varsigma}]$
gives $\alpha$ for ($\mathbf{k}_{0}\varsigma_{0}$)-mode and 0 for
other modes. Under the rotating-wave approximation, the above interaction
becomes 
\begin{equation}
\tilde{H}_{\text{\textsc{sb}}}^{(0)}(t)\simeq\sum_{\mathbf{k},\varsigma}g_{\mathbf{k}\varsigma}\,\hat{\sigma}^{+}\,\delta\hat{a}_{\mathbf{k}\varsigma}\,e^{i(\Omega-\omega_{\mathbf{k}})t}+\text{h.c.}
\end{equation}
 where $g_{\mathbf{k}\varsigma}:=-i(\vec{\wp}\cdot\hat{\mathrm{e}}_{\mathbf{k}\varsigma})\sqrt{\hbar\omega_{\mathbf{k}}/2\epsilon_{0}V}\,e^{i\mathbf{k}\cdot\mathbf{x}}$.

In the interaction picture, the dynamics of the system-bath state
$\tilde{\boldsymbol{\rho}}_{\text{\textsc{sb}}}(t)$ is governed by
the von Neumann equation, 
\begin{eqnarray}
\partial_{t}\tilde{\boldsymbol{\rho}}_{\text{\textsc{sb}}}(t) & = & \frac{i}{\hbar}[\tilde{\boldsymbol{\rho}}_{\text{\textsc{sb}}}(t),\,\tilde{V}_{\alpha}(t)]+\frac{i}{\hbar}[\tilde{\boldsymbol{\rho}}_{\text{\textsc{sb}}}(t),\,\tilde{H}_{\text{\textsc{sb}}}^{(0)}(t)],\nonumber \\
\tilde{\boldsymbol{\rho}}_{\text{\textsc{sb}}}(t) & = & \tilde{\boldsymbol{\rho}}_{\text{\textsc{sb}}}(0)+\frac{i}{\hbar}\int_{0}^{t}ds\,[\tilde{\boldsymbol{\rho}}_{\text{\textsc{sb}}}(s),\,\tilde{V}_{\alpha}(s)+\tilde{H}_{\text{\textsc{sb}}}^{(0)}(s)].
\end{eqnarray}
We put the above integral solution of $\tilde{\boldsymbol{\rho}}_{\text{\textsc{sb}}}(t)$
back into the second term of the von Neumann equation, which gives
\begin{eqnarray}
\partial_{t}\tilde{\boldsymbol{\rho}}_{\text{\textsc{sb}}}(t) & = & \frac{i}{\hbar}[\tilde{\boldsymbol{\rho}}_{\text{\textsc{sb}}}(t),\,\tilde{V}_{\alpha}(t)]+\frac{i}{\hbar}[\tilde{\boldsymbol{\rho}}_{\text{\textsc{sb}}}(0),\,\tilde{H}_{\text{\textsc{sb}}}^{(0)}(t)]\nonumber \\
 & - & \frac{1}{\hbar^{2}}\int_{0}^{t}ds\,\Big[[\tilde{\boldsymbol{\rho}}_{\text{\textsc{sb}}}(s),\,\tilde{V}_{\alpha}(s)+\tilde{H}_{\text{\textsc{sb}}}^{(0)}(s)],\,\tilde{H}_{\text{\textsc{sb}}}^{(0)}(t)\Big].\label{eq:Iteration}
\end{eqnarray}
Now we apply the Born approximation $\tilde{\boldsymbol{\rho}}_{\text{\textsc{sb}}}(s)\simeq\tilde{\rho}_{\text{\textsc{s}}}(s)\otimes\hat{\boldsymbol{\rho}}_{\mathnormal{\textsc{b}}}^{(\alpha)_{\mathbf{k}_{0}\varsigma_{0}}}(0)$,
and trace out the bath degree of freedom. Since $\langle\delta\hat{a}_{\mathbf{k}\varsigma}\rangle=\langle\delta\hat{a}_{\mathbf{k}\varsigma}^{\dagger}\rangle=0$
under the bath state $\hat{\boldsymbol{\rho}}_{\mathnormal{\textsc{b}}}^{(\alpha)_{\mathbf{k}_{0}\varsigma_{0}}}(0)$,
equation (\ref{eq:Iteration}) further gives 
\begin{eqnarray}
 &  & \partial_{t}\tilde{\rho}_{\text{\textsc{s}}}\simeq\frac{i}{\hbar}[\tilde{\rho}_{\text{\textsc{s}}}(t),\,\tilde{V}_{\alpha}(t)]\nonumber \\
 &  & -\frac{1}{\hbar^{2}}\int_{0}^{t}ds\,\mathrm{Tr}_{\text{\textsc{b}}}\Big[[\tilde{\rho}_{\text{\textsc{s}}}(t-s)\otimes\hat{\boldsymbol{\rho}}_{\mathnormal{\textsc{b}}}^{(\alpha)_{\mathbf{k}_{0}\varsigma_{0}}}(0),\,\tilde{H}_{\text{\textsc{sb}}}^{(0)}(t-s)],\,\tilde{H}_{\text{\textsc{sb}}}^{(0)}(t)\Big].
\end{eqnarray}
 Then we assume the convolution kernel, which comes from the time
correlation function of the EM field, decays so fast that only the
accumulation around $\tilde{\rho}_{\text{\textsc{s}}}(t-s\simeq t)$
dominates in the integral. Thus, we can extend the above time integral
to be $t\rightarrow\infty$ (Markovian approximation), and obtain
\begin{eqnarray}
 &  & \partial_{t}\tilde{\rho}_{\text{\textsc{s}}}\simeq\frac{i}{\hbar}[\tilde{\rho}_{\text{\textsc{s}}}(t),\,\tilde{V}_{\alpha}(t)]\nonumber \\
 &  & -\frac{1}{\hbar^{2}}\int_{0}^{\infty}ds\,\mathrm{Tr}_{\text{\textsc{b}}}\Big[[\tilde{\rho}_{\text{\textsc{s}}}(t)\otimes\hat{\boldsymbol{\rho}}_{\mathnormal{\textsc{b}}}^{(\alpha)_{\mathbf{k}_{0}\varsigma_{0}}}(0),\,\tilde{H}_{\text{\textsc{sb}}}^{(0)}(t-s)],\,\tilde{H}_{\text{\textsc{sb}}}^{(0)}(t)\Big].\label{eq:ME-0}
\end{eqnarray}

The master equation can be obtained after taking the trace expectation
and time integral. Notice that, when taking the average on the bath
state $\hat{\boldsymbol{\rho}}_{\mathnormal{\textsc{b}}}^{(\alpha)_{\mathbf{k}_{0}\varsigma_{0}}}(0)$,
the bath operators $\delta\hat{a}_{\mathbf{k}\varsigma}^{\dagger}$
in $\tilde{H}_{\text{\textsc{sb}}}^{(0)}(t)$ satisfy the following
relations, 
\begin{eqnarray}
\langle\delta\hat{a}_{\mathbf{k}\varsigma}^{\dagger}\,\delta\hat{a}_{\mathbf{k}'\varsigma'}\rangle & = & 0,\quad\langle\delta\hat{a}_{\mathbf{k}\varsigma}\,\delta\hat{a}_{\mathbf{k}'\varsigma'}^{\dagger}\rangle=\delta_{\mathbf{k}\mathbf{k}'}\delta_{\varsigma\varsigma'},\nonumber \\
\langle\delta\hat{a}_{\mathbf{k}\varsigma}\,\delta\hat{a}_{\mathbf{k}'\varsigma'}\rangle & = & \langle\delta\hat{a}_{\mathbf{k}\varsigma}^{\dagger}\,\delta\hat{a}_{\mathbf{k}'\varsigma'}^{\dagger}\rangle=0.
\end{eqnarray}

Here we present the calculation of one term in equation (\ref{eq:ME-0}):
\begin{eqnarray}
 &  & -\frac{1}{\hbar^{2}}\int_{0}^{\infty}ds\,\mathrm{Tr}_{\text{\textsc{b}}}\Big[\tilde{\rho}_{\text{\textsc{s}}}(t)\otimes\hat{\boldsymbol{\rho}}_{\mathnormal{\textsc{b}}}^{(\alpha)_{\mathbf{k}_{0}\varsigma_{0}}}(0)\cdot\Big(\sum_{\mathbf{k}\varsigma}g_{\mathbf{k}\varsigma}\hat{\sigma}^{+}\delta\hat{a}_{\mathbf{k}\varsigma}\,e^{i(\Omega-\omega_{\mathbf{k}})(t-s)}\Big)\nonumber \\
 &  & \qquad\cdot\Big(\sum_{\mathbf{k}'\varsigma'}g_{\mathbf{k}'\varsigma'}^{*}\hat{\sigma}^{-}\delta\hat{a}_{\mathbf{k}'\varsigma'}^{\dagger}\,e^{-i(\Omega-\omega_{\mathbf{k}'})t}\Big)\Big]\nonumber \\
 &  & =-\tilde{\rho}_{\text{\textsc{s}}}\hat{\sigma}^{+}\hat{\sigma}^{-}\,\sum_{\mathbf{k}\varsigma}\frac{|g_{\mathbf{k}\varsigma}|^{2}}{\hbar^{2}}\int_{0}^{\infty}ds\,\langle\delta\hat{a}_{\mathbf{k}\varsigma}\delta\hat{a}_{\mathbf{k}\varsigma}^{\dagger}\rangle e^{-i(\Omega-\omega_{\mathbf{k}})s}\nonumber \\
 &  & =-\tilde{\rho}_{\text{\textsc{s}}}\hat{\sigma}^{+}\hat{\sigma}^{-}\int_{0}^{\infty}\frac{d\omega}{2\pi}\,\Gamma(\omega)\int_{0}^{\infty}ds\,e^{-i(\Omega-\omega)s}\nonumber \\
 &  & =-\tilde{\rho}_{\text{\textsc{s}}}\hat{\sigma}^{+}\hat{\sigma}^{-}\int_{0}^{\infty}\frac{d\omega}{2\pi}\,\Gamma(\omega)[\pi\delta(\Omega-\omega)-i\mathbf{P}\frac{1}{\Omega-\omega}]\nonumber \\
 &  & \simeq-\frac{1}{2}\Gamma(\Omega)\,\tilde{\rho}_{\text{\textsc{s}}}\hat{\sigma}^{+}\hat{\sigma}^{-}.
\end{eqnarray}
Here the principal integral is omitted, and $\Gamma(\omega):=\frac{2\pi}{\hbar^{2}}\sum_{\mathbf{k}\varsigma}|g_{\mathbf{k}\varsigma}|^{2}\delta(\omega-\omega_{\mathbf{k}\varsigma})$
is the coupling spectral density. 

Notice that, when considering the spontaneous emission of the TLS
in the vacuum field (without the driving light), the coupling spectral
density $\Gamma(\omega)$ is exactly the same with the one used here.
Finally, the master equation is obtained as 
\begin{equation}
\partial_{t}\tilde{\rho}_{\text{\textsc{s}}}=\frac{i}{\hbar}[\tilde{\rho}_{\text{\textsc{s}}},\tilde{V}_{\alpha}(t)]+\kappa\big(\hat{\sigma}^{-}\tilde{\rho}_{\text{s}}\hat{\sigma}^{+}-\frac{1}{2}\{\hat{\sigma}^{+}\hat{\sigma}^{-},\,\tilde{\rho}_{\text{\textsc{s}}}\}\big).
\end{equation}
The first term just has the form of quasi-classical driving widely
adopted in literature, and second term describes the spontaneous emission
with $\kappa:=\Gamma(\Omega)$ as the decay rate.

\section{Photoelectric current \label{sec:Photoelectric-current}}

Based on the master equation equation (\ref{eq:QD-ME}) in the main
text, we obtain the following the equations of motion for the observables
$\hat{\text{\textsc{n}}}_{a}=\hat{a}^{\dagger}\hat{a}$, $\hat{\text{\textsc{n}}}_{b}=\hat{b}^{\dagger}\hat{b}$,
and $\hat{\tau}^{+}=\hat{a}^{\dagger}\hat{b}=(\hat{\tau}^{-})^{\dagger}$,
\begin{eqnarray}
\fl\partial_{t}\langle\hat{\text{\textsc{n}}}_{a}\rangle=\big(\xi_{0}\alpha\langle\hat{\tau}^{+}\rangle+\xi_{0}^{*}\alpha^{*}\langle\hat{\tau}^{-}\rangle\big)-\gamma_{a}[(1-\bar{\mathtt{n}}_{a})\langle\hat{\text{\textsc{n}}}_{a}\rangle-\bar{\mathtt{n}}_{a}(1-\langle\hat{\text{\textsc{n}}}_{a}\rangle)]-\kappa\langle\hat{\text{\textsc{n}}}_{a}\rangle,\nonumber \\
\fl\partial_{t}\langle\hat{\text{\textsc{n}}}_{b}\rangle=-\big(\xi_{0}\alpha\langle\hat{\tau}^{+}\rangle+\xi_{0}^{*}\alpha^{*}\langle\hat{\tau}^{-}\rangle\big)-\gamma_{b}[(1-\bar{\mathtt{n}}_{b})\langle\hat{\text{\textsc{n}}}_{b}\rangle-\bar{\mathtt{n}}_{b}(1-\langle\hat{\text{\textsc{n}}}_{b}\rangle)]+\kappa\langle\hat{\text{\textsc{n}}}_{a}\rangle,\nonumber \\
\fl\partial_{t}\langle\hat{\tau}^{+}\rangle=-\xi_{0}^{*}\alpha^{*}\big(\langle\hat{\text{\textsc{n}}}_{a}\rangle-\langle\hat{\text{\textsc{n}}}_{b}\rangle\big)-\frac{1}{2}(\gamma_{a}+\gamma_{b}+\kappa)\langle\hat{\tau}^{+}\rangle,\nonumber \\
\fl\partial_{t}\langle\hat{\tau}^{-}\rangle=-\xi_{0}\alpha\big(\langle\hat{\text{\textsc{n}}}_{a}\rangle-\langle\hat{\text{\textsc{n}}}_{b}\rangle\big)-\frac{1}{2}(\gamma_{a}+\gamma_{b}+\kappa)\langle\hat{\tau}^{-}\rangle.
\end{eqnarray}
In the steady state $t\rightarrow\infty$, the time-derivatives all
give zero, and the above algebra equations give the steady state as
\begin{eqnarray}
\fl\langle\hat{\text{\textsc{n}}}_{a}\rangle=\frac{4|\alpha|^{2}|\xi_{0}|^{2}(\gamma_{a}\bar{\mathtt{n}}_{a}+\gamma_{b}\bar{\mathtt{n}}_{b})+\gamma_{a}\gamma_{b}(\gamma_{a}+\gamma_{b}+\kappa)\bar{\mathtt{n}}_{a}}{4|\alpha|^{2}|\xi_{0}|^{2}(\gamma_{a}+\gamma_{b})+\gamma_{b}(\gamma_{a}+\kappa)(\gamma_{a}+\gamma_{b}+\kappa)},\nonumber \\
\fl\langle\hat{\text{\textsc{n}}}_{b}\rangle=\frac{4|\alpha|^{2}|\xi_{0}|^{2}(\gamma_{a}\bar{\mathtt{n}}_{a}+\gamma_{b}\bar{\mathtt{n}}_{b})+(\gamma_{a}+\gamma_{b}+\kappa)[\kappa\gamma_{a}\bar{\mathtt{n}}_{a}+(\kappa+\gamma_{a})\gamma_{b}\bar{\mathtt{n}}_{b}]}{4|\alpha|^{2}|\xi_{0}|^{2}(\gamma_{a}+\gamma_{b})+\gamma_{b}(\gamma_{a}+\kappa)(\gamma_{a}+\gamma_{b}+\kappa)},\\
\fl\langle\hat{\tau}^{+}\rangle=\langle\hat{\tau}^{-}\rangle^{*}=\frac{2\xi_{0}^{*}\alpha^{*}[\gamma_{a}\gamma_{b}(\bar{\mathtt{n}}_{b}-\bar{\mathtt{n}}_{a})+\kappa(\gamma_{a}\bar{\mathtt{n}}_{a}+\gamma_{b}\bar{\mathtt{n}}_{b})]}{4|\alpha|^{2}|\xi_{0}|^{2}(\gamma_{a}+\gamma_{b})+\gamma_{b}(\gamma_{a}+\kappa)(\gamma_{a}+\gamma_{b}+\kappa)}.\nonumber 
\end{eqnarray}
Then the electron current flowing to the right electron lead is given
by 
\begin{eqnarray}
J_{\text{\textsc{r}}} & = & -\mathrm{tr}\Big\{{\cal L}_{a}[\tilde{\rho}_{\text{\textsc{s}}}]\cdot\hat{\text{\textsc{n}}}_{a}\Big\}=\gamma_{a}[(1-\bar{\mathtt{n}}_{a})\langle\hat{\text{\textsc{n}}}_{a}\rangle-\bar{\mathtt{n}}_{a}(1-\langle\hat{\text{\textsc{n}}}_{a}\rangle)]\nonumber \\
 & = & \frac{4|\alpha|^{2}\,|\xi_{0}|^{2}\gamma_{a}\gamma_{b}(\bar{\mathtt{n}}_{b}-\bar{\mathtt{n}}_{a})-\kappa\gamma_{a}\gamma_{b}(\gamma_{a}+\gamma_{b}+\kappa)\bar{\mathtt{n}}_{a}}{4|\alpha|^{2}|\xi_{0}|^{2}(\gamma_{a}+\gamma_{b})+\gamma_{b}(\gamma_{a}+\kappa)(\gamma_{a}+\gamma_{b}+\kappa)}.\label{eq:J-R}
\end{eqnarray}
 Taking $\gamma_{a}=\gamma_{b}:=\gamma$, $\kappa=0$, $\bar{\mathtt{n}}_{a}=0$,
$\bar{\mathtt{n}}_{b}=1$, it gives the result (\ref{eq:J-alpha})
in the main text. 

Notice that, the second term started with $(-\kappa)$ in the above
numerator indeed indicates the electron tunneling from level-\emph{a}
to level-\emph{b} under the mediation of spontaneous emission, and
it still exists when there is no driving light ($\alpha\rightarrow0$).
In this paper, we neglect this effect since the spontaneous rate $\kappa$
is usually much smaller than the tunneling rates $\gamma_{a,b}$. 

\section{General input photon statistics \label{sec:Generic-input-photon}}

Here we show how to calculate the photoelectric current when the input
light is not a coherent state but has a general photon statistics.
Generally, the \emph{P} function average of equation (\ref{eq:J-alpha})
in the main text gives the photoelectric current. But for many nonclassical
light states, their \emph{P} functions are highly singular and sometimes
not easy to be given directly. Thus here we provide another method
to calculate this current. Notice that the \emph{P} function average
is also equivalent as the normal-order expectation on the quantum
state $\rho=\int d^{2}\alpha\,P(\alpha)|\alpha\rangle\langle\alpha|$,
thus we have (denoting $\tilde{\gamma}_{\xi}:=\gamma/|\xi_{0}|$)
\begin{eqnarray}
\overline{J} & = & \int d^{2}\alpha\,P(\alpha)\,\frac{2|\xi_{0}|^{2}\gamma\,|\alpha|^{2}}{4|\xi_{0}|^{2}\,|\alpha|^{2}+\gamma^{2}}=\frac{\gamma}{2}-\frac{\gamma}{2}\big\langle:\frac{\tilde{\gamma}_{\xi}^{2}}{4\hat{a}^{\dagger}\hat{a}+\tilde{\gamma}_{\xi}^{2}}:\big\rangle\nonumber \\
 & = & \frac{\gamma}{2}\Big(1-\tilde{\gamma}_{\xi}^{2}\big\langle:\int_{0}^{\infty}ds\,e^{-s(4\hat{a}^{\dagger}\hat{a}+\tilde{\gamma}_{\xi}^{2})}:\big\rangle\Big).
\end{eqnarray}
Here $\langle:f(\hat{a},\hat{a}^{\dagger}):\rangle$ means the normal-order
expectation, and the second line is the Widder transform which turns
the operator fraction into an exponential integral. Thus, for an arbitrary
quantum state $\rho=\sum_{mn}\rho_{mn}|m\rangle\langle n|$, we have
\begin{eqnarray}
\langle:e^{-4s\hat{a}^{\dagger}\hat{a}}:\rangle & = & \sum_{k=0}^{\infty}\frac{(-4s)^{k}}{k!}\langle(\hat{a}^{\dagger})^{k}\hat{a}^{k}\rangle=\sum_{m,n=0}^{\infty}\sum_{k=0}^{\infty}\rho_{mn}\cdot\frac{(-4s)^{k}}{k!}\langle n|(\hat{a}^{\dagger})^{k}\hat{a}^{k}|m\rangle\nonumber \\
 & = & \sum_{n=0}^{\infty}\sum_{k=0}^{n}\rho_{nn}\cdot\frac{(-4s)^{k}n!}{k!(n-k)!}=\sum_{n}\rho_{nn}(1-4s)^{n}.
\end{eqnarray}
Indeed here $\rho_{nn}:=P_{n}$ is just the photon statistics of the
input light state, and the above photoelectric current becomes 
\begin{equation}
\overline{J}=\frac{\gamma}{2}\Big(1-\tilde{\gamma}_{\xi}^{2}\int_{0}^{\infty}ds\,e^{-\tilde{\gamma}_{\xi}^{2}s}\big[\sum_{n}P_{n}(1-4s)^{n}\big]\Big).\label{eq:current-Pn}
\end{equation}

For example, considering the coherent state $|\alpha\rangle$ as the
input light, which has $P_{n}=e^{-|\alpha|^{2}}|\alpha|^{2n}/n!$,
then equation (\ref{eq:current-Pn}) gives the photoelectric current
by 
\begin{eqnarray}
\overline{J} & = & \frac{\gamma}{2}\Big(1-\tilde{\gamma}_{\xi}^{2}\int_{0}^{\infty}ds\,e^{-\tilde{\gamma}_{\xi}^{2}s}\big[\sum_{n}e^{-|\alpha|^{2}}\frac{|\alpha|^{2n}(1-4s)^{n}}{n!}\big]\Big)\nonumber \\
 & = & \frac{\gamma}{2}\Big(1-\tilde{\gamma}_{\xi}^{2}\int_{0}^{\infty}ds\,e^{-\tilde{\gamma}_{\xi}^{2}s}\,e^{-4|\alpha|^{2}s}\Big)=\frac{\gamma}{2}\Big[1-\frac{\tilde{\gamma}_{\xi}^{2}}{4|\alpha|^{2}+\tilde{\gamma}_{\xi}^{2}}\Big].
\end{eqnarray}
which just returns the result $J(\alpha)$ {[}equation (\ref{eq:J-alpha})
in the main text{]}. If we consider the input light is the thermal
state $P_{n}=\frac{1}{\bar{n}+1}\big[\frac{\bar{n}}{\bar{n}+1}\big]^{n}$,
the above equation (\ref{eq:current-Pn}) also gives the same result
as equation (\ref{eq:J-th}) in the main text. If the input light
has a sub-Poisson statistics $P_{n}=[I_{0}(2\sqrt{\lambda})]^{-1}\,\lambda^{n}/(n!)^{2}$,
the above equation (\ref{eq:current-Pn}) gives the result (\ref{eq:J_sub})
in the main text.

\section{Proof for the classical upper bound \label{sec:Proof-for-the}}

Here we are going to show, among all the classical light states, under
the same mean photon number, the Poisson light generates the largest
photoelectric current.

We have seen that, for different input light states, the photoelectric
currents are given by 
\begin{equation}
\overline{J}/\gamma=\int d^{2}\alpha\,P(\alpha,\alpha^{*})\,\frac{2|\alpha|^{2}}{4|\alpha|^{2}+\tilde{\gamma}_{\xi}^{2}},
\end{equation}
where $P(\alpha,\alpha^{*})$ is the \emph{P} function of the input
light state. Therefore, the classical upper bound for the photoelectric
current can be obtained by finding the variational extremum of this
integral under three constraints: (1) classical light state $P(\alpha,\alpha^{*})\ge0$,
(2) normalization $\int d^{2}\alpha\,P(\alpha)=1$, (3) fixed mean
photon number $\int d^{2}\alpha\,|\alpha|^{2}P(\alpha)=\overline{n}$.

Since the \emph{P} function of classical light states must be positive,
and no more singular than the $\delta$-function, we introduce $[p(\alpha,\alpha^{*})]^{2}\equiv P(\alpha,\alpha^{*})\ge0$
to handle the positivity constraint. Then the above extremum problem
can be done with the help of Lagrangian multipliers ($\lambda_{1,2}$),
namely, 
\begin{eqnarray}
S & := & \int d^{2}\alpha\,\frac{2|\alpha|^{2}}{4|\alpha|^{2}+\tilde{\gamma}_{\xi}^{2}}[p(\alpha)]^{2}-\lambda_{1}\big\{\int d^{2}\alpha\,[p(\alpha)]^{2}-1\big\}\nonumber \\
 &  & \qquad-\lambda_{2}\big\{\int d^{2}\alpha\,|\alpha|^{2}[p(\alpha)]^{2}-\overline{n}\big\},\nonumber \\
\delta S & = & \int d^{2}\alpha\,\Big\{\big[\frac{2|\alpha|^{2}}{4|\alpha|^{2}+\tilde{\gamma}_{\xi}^{2}}-\lambda_{1}-\lambda_{2}|\alpha|^{2}\big]\,2p(\alpha)\Big\}\,\delta p(\alpha).
\end{eqnarray}

To make sure the extremum condition $\delta S\equiv0$ holds for any
variance $\delta p(\alpha)$, the term in the above curly bracket
must be zero, and thus $P(\alpha,\alpha^{*})$ must satisfy the following
relation, \begin{equation}
P(\alpha,\alpha^{*})=\cases{[p(\alpha)]^{2}\neq0, & when $\frac{2|\alpha|^{2}}{4|\alpha|^{2}+\tilde{\gamma}_{\xi}^{2}}-\lambda_{1}-\lambda_{2}|\alpha|^{2}=0$ \\
{}[p(\alpha)]^{2}=0, & for other $\alpha$ }
\end{equation} That means $P(\alpha,\alpha^{*})$ is zero unless $|\alpha|^{2}$
equals to a certain value. Then together with the above constraints
(2, 3), $P(\alpha,\alpha^{*})$ must have the following form, 
\begin{equation}
P(\alpha,\alpha^{*})=\int_{0}^{2\pi}d\phi\,f(\phi)\delta^{(2)}(\alpha-\sqrt{\bar{n}}\,e^{i\phi}),
\end{equation}
where $f(\phi)$ is an arbitrary function satisfying $f(\phi)>0$
and $\int_{0}^{2\pi}d\phi\,f(\phi)=1$. That means, the light state
$\rho=\int d^{2}\alpha\,P(\alpha)|\alpha\rangle\langle\alpha|$ is
indeed a mixture of many coherent states $\big|\alpha=\sqrt{\bar{n}}e^{i\phi}\big\rangle$,
which have the same mean photon number $|\alpha|^{2}=\overline{n}$
but different phases $\phi$. Clearly, all such states have the same
Poisson statistics, and generates the photoelectric current as equation
(\ref{eq:J-alpha}) in the main text. 

Therefore, when the mean photon number $\overline{n}$ is fixed, the
Poisson input generates the largest photoelectric current among all
classical light states. For many nonclassical states, the \emph{P}
functions are highly singular {[} such as containing high-order derivatives
of the $\delta$-function, e.g., the Fock states have $P_{|n\rangle}(\alpha)=(e^{|\alpha|^{2}}/n!)\,\partial_{\alpha}^{n}\partial_{\alpha^{*}}^{n}\delta(\alpha)$
{]}, thus the above variational method does not apply well in the
functional space of nonclassical states.

\providecommand{\newblock}{}

\end{document}